\documentclass[preprint]{aastex}

\def\la{\mathrel{\hbox{\rlap{\hbox{\lower4pt\hbox{$\sim$}}}\hbox{$<$}}}}
\def\ga{\mathrel{\hbox{\rlap{\hbox{\lower4pt\hbox{$\sim$}}}\hbox{$>$}}}}

\def\etal{et\ al.}
\def\CIV{\ion{C}{4}}
\def\MgII{\ion{Mg}{2}}

\def\hMpc{$h^{-1}$~Mpc}
\def\kms{km~s$^{-1}$}

\slugcomment{Submitted to {\it The Astrophysical Journal}}

\shorttitle{Clustering of Absorbers}
\shortauthors{Loh, Quashnock, \& Stein}

\begin{document}

\title{A Measurement of the Three--Dimensional Clustering\\
of \CIV\ Absorption--Line Systems on Scales of 5 to 300 \hMpc}

\author{Ji-Meng Loh}
\affil{Department of Statistics, University of Chicago, Chicago, IL 60637}
\email{loh@galton.uchicago.edu}

\author{Jean M. Quashnock\altaffilmark{1}}
\affil{Department of Physics, Carthage College, Kenosha, WI 53140}
\email{jmq@carthage.edu}

\and

\author{Michael L. Stein}
\affil{Department of Statistics, University of Chicago, Chicago, IL 60637}
\email{stein@galton.uchicago.edu}

\altaffiltext{1}{Visiting Scholar, Department of Astronomy \& Astrophysics,
University of Chicago, Chicago, IL 60637}

\begin{abstract}
We examine the three--dimensional clustering of \CIV\ absorption--line systems,
using an extensive catalog of QSO heavy--element absorbers drawn from the literature.
We measure clustering by a volume--weighted integral of the correlation function
called the reduced second--moment measure,
and include information from both {\em along} and {\em across} QSO lines of sight,
thus enabling a full determination of the three--dimensional
clustering of absorbers, as well as a comparison of line--
and cross--line--of--sight clustering properties.
Here we present the three--dimensional reduced second--moment estimator for a
three--dimensional point process probed by one--dimensional lines of sight,
and apply our algorithm to a sample of 345 \CIV\ absorbers
with median redshift $\langle z\rangle = 2.2$, drawn from the spectra of 276 QSOs.
We confirm the existence of significant clustering on comoving
scales up to 100 \hMpc\ ($q_0 = 0.5$),
and find that the additional {\em cross}--line--of--sight information strengthens the evidence
for clustering on scales from 100 \hMpc\ to 150 \hMpc.
There is no evidence of absorber clustering along or across lines of sight for
scales from 150 \hMpc\ to 300 \hMpc.
We show that with a 300--times larger catalog, such as that to be compiled
by the Sloan Digital Sky Survey (100,000 QSOs),
use of the full three--dimensional estimator and cross--line--of--sight information
will {\em substantially} increase clustering sensitivity.
We find that standard errors are reduced by a factor 2 to 20 on scales of 30 to 200 \hMpc,
{\em in addition} to the factor of $\sqrt{300}$ reduction from the larger sample size,
effectively increasing the sample size by an extra factor of 4 to 400 at large distances.
\end{abstract}

\keywords{cosmology:observations --- intergalactic medium --- large--scale
structure of universe --- methods:statistical --- quasars:absorption lines}

\section{Introduction}

In a previous series of investigations \citep{VQYY96,Quash96,Quash98},
the clustering properties of \CIV\ and \MgII\ absorbers have been
investigated, using an extensive catalog of heavy--element absorption--line
systems drawn from the literature.\footnote{Contact D. E. Vanden Berk
(danvb@fnal.gov) for the latest version of the catalog;
see \citet{Y91} for an earlier version.}
These authors used a one--dimensional correlation analysis ---
one confined to pairs of absorbers along the {\em same} QSO line of sight ---
and found  evidence for strong and evolving clustering on small scales (1--16 \hMpc),
as well as for superclustering on scales as large as 50--100 \hMpc.
Together, these investigations suggest that these strong absorbers,
with median rest equivalent width $\langle W\rangle=0.4$ \AA\ for \CIV\
\citep{Quash98},
are biased tracers of the higher density regions of space, and that agglomerations
of absorbers along a line of sight are indicators of clusters and superclusters.

Recently, \citet{Quash99} used a new measure of clustering, called the
reduced second moment measure or $K(r)$ \citep{Ripley88,Badd98},
which directly measures the mean overdensity of absorbers on scales $\la r$.
While closely related to other second--order measures of clustering,
such as the correlation function or the
power spectrum, the reduced second moment measure nevertheless
has a number of advantageous statistical properties, and has recently been studied
by astrophysicists \citep{Mart98,Quash99,Stein00}.
\citet{Quash99} found significant evidence for clustering of \CIV\
absorbers on scales of 20--100 \hMpc, with marginal evidence
on 100--200 \hMpc\ scales, again suggesting that the absorbers show
superclustering much like what is seen locally in the distribution of galaxies.

However, their analysis also was confined to pairs of
absorbers along the same line of sight,
and hence did not include {\em cross}--line--of--sight information valuable for
a full determination of the three--dimensional clustering properties of absorbers,
which we are ultimately interested in finding.
Indeed, \citet{RY99} have claimed that there is evidence of some significant contamination
of true intervening systems along the line of sight by absorbers
that are actually physically
associated with the QSO, and that such contamination may extend to relative velocities
as great as 75000 \kms. This means that such a contamination could be present as well in the
superclustering signal found by \citet{Quash99}.
In addition, on smaller scales of a few megaparsecs,
\citet{Crotts97} have claimed that the clustering of \CIV\ systems {\em across}
adjacent lines of sight is significantly weaker than {\em along} a line of sight
and have questioned whether the correlations are due to
velocity dispersion in associated systems rather than intrinsic spatial clustering.
The number of adjacent lines analyzed by \citet{Crotts97} was small, however.

It is thus essential to
study correlations both {\em along} and {\em across} QSO lines of sight,
to enable a complete determination of the three--dimensional
clustering of absorbers, as well as a comparison of line--
and cross--line--of--sight clustering properties
(and ultimately statistically discriminate between intrinsic and intervening absorbers).
What is required is a method that measures spatial clustering of absorbers ---
which are nonetheless confined to QSO lines of sight ---
and can contrast clustering along and across lines of sight.

Here we present the three--dimensional estimator for the reduced second moment measure of a
three--dimensional point process probed by one--dimensional lines of sight.
We apply our algorithm to a sample of 345 \CIV\ absorbers
with median redshift $\langle z\rangle = 2.2$, drawn from the spectra of 276 QSOs
in the aforementioned catalog of Vanden Berk \etal,
with the goal of determining the clustering of absorbers on very large scales
and seeing if the signal found by \citet{Quash99} remains when including
cross--line--of--sight information, or, more compellingly, using only
cross--line--of--sight information.

The outline of the paper is as follows:
In \S 2 we define the
three--dimensional reduced second moment measure, compare it
to its one--dimensional analog, and outline our method of estimation.
In \S 3 we apply our methodology to the above sample and present our results.
In \S 4 we interpret the results and discuss how
the full three--dimensional estimator and cross--line--of--sight information
will improve our ability to measure clustering with the absorber
sample from the Sloan Digital Sky Survey (100,000 QSOs).
Finally, we conclude in \S 5 and present the details of the
three--dimensional reduced second moment estimator in the Appendix.

\section{The Reduced Second Moment Measure}

Here we assume that the clustering of absorbers is both statistically homogeneous
and stationary (does not depend on cosmic epoch or redshift $z$)
when examined in comoving coordinates. The latter assumption is likely not to be
strictly true, since growth of the correlation function with decreasing
redshift has been detected, at least on smaller scales
of 1--16 \hMpc \citep{Quash98}. Nevertheless, our results here can be thought of as averages
for the absorber sample as a whole, which has a characteristic redshift
given by the median $\langle z\rangle = 2.2$. It is possible to extend our treatment
and examine the evolution of the clustering with redshift (see \S 4 below
regarding the Sloan Digital Sky Survey), but we have not done so here,
largely because of the limited size of our sample.
We follow the usual convention and take the Hubble constant, $H_0$, to be
100~$h$~\kms~Mpc$^{-1}$ and take $q_0=0.5$ and $\Lambda=0$.

\subsection{Definitions}

We treat the distribution of
absorbers as a point process in {\em three}--dimensional space
rather than as a {\em one}--dimensional process on the lines of sight,
and use the reduced second moment to describe the three--dimensional clustering.
Otherwise, we closely follow the treatment given in \citet{Quash99} ---
where one--dimensional clustering is described ---
and use the same symbol $K$ for the three--dimensional reduced second moment.

Let $\lambda$ be the mean number of absorbers per unit comoving volume.
The absorbers have some physical size, perhaps of order 100 kpc or so \citep{Church96},
so that there is a finite probability of intersection between an absorber and the
QSO lines of sight. For simplicity, we assume the absorbers are balls
of identical radius $d$. Our treatment is accurate on scales of
a few Mpc and greater,
which are much larger than the physical size of the absorbers.
The clustering we seek to measure is that of the {\em centers} of the absorbers.

Neither our methods nor results depend on specifying $d$ (see Appendix).
Since absorbers do vary in size, mass and column density,
and their clustering likely depends on these quantities \citep{Cristi97,Dodori98}, 
our results cannot be directly interpreted as a quantitative
measure of the clustering of mass.
Nevertheless, we expect our results to be qualitatively correct to the
extent that if our estimates of $K$ show evidence of clustering on some
spatial scale, matter should also show clustering on this same scale,
especially if the latter is quite large.
Furthermore, our results are neither more nor less dependent on the assumption
of equal absorber size as those in the one--dimensional analyses presented in
\citet{Quash98} or \cite{Quash99};
rather, it is just harder to ignore the fact that absorbers must have a
finite cross section and volume 
when doing a three--dimensional analysis based on intersections
of absorbers with lines of sight.

The {\em reduced second moment measure}, $K(r)$, is the
conditional expectation, or average ---
given that there is an absorber center at $x$ ---
of the number of absorbers (other than the one at $x$ itself),
$N(x,r)$, whose centers are within a comoving distance $r$ of $x$,
normalized by $\lambda$:
\begin{equation}
K(r)=\frac{1}{\lambda} \, E\left[N(x,r) \mid
{\rm absorber\; at}\; x\right] \;  .
\end{equation}
Because of our assumption of homogeneity,
the expected number of absorbers in equation~(1) does not depend on $x$.
With $q_0=0.5$ and $\Lambda=0$, the comoving distance $r$ between two
absorbers at redshifts $z_1$ and $z_2$ is
$ r = 2c/ H_0 \times \left|{1/\sqrt{1+z_1}} - {1/\sqrt{1+z_2}}\right| $.

In terms of the two--point correlation function $\xi(r)$
\citep{Peeb80,Peeb93}, the reduced second moment measure is given by
\begin{equation}
K(r) = 4\pi \int_{0}^{r} u^2\, du \, \left[1+\xi(u)\right]\; .
\end{equation}
If no correlations are present, then $K(r)= \case{4}{3}\pi r^3$.
Simply put, in this case the number of surrounding absorber centers within
distance $r$ of $x$ would not depend on the
fact that there is an absorber center at $x$, and
would simply be equal to $\case{4}{3}\pi r^3 \lambda$. The quantity
$K(r)/(\case{4}{3}\pi r^3) \equiv 1+ \rho(r)$ is then a measure of the relative mean density
of absorbers around other absorbers, averaged over scales less than $r$.
Thus the reduced second moment measure is essentially a volume--weighted
integral of the correlation function, whereas in the one--dimensional treatment
of \citet{Quash99} it is a distance--weighted integral.
Even if one does not use across--line--of--sight information, it is arguably
more natural to study the three-dimensional reduced second moment measure.

The relative mean {\em over--density}, $\rho(r)$,
can be written in terms of the power spectrum,
$P(k)$, the Fourier transform of the correlation function $\xi(r)$,
or equivalently, in terms of the dimensionless power per logarithmic
wavenumber, $\Delta^2(k)\equiv k^3P(k)/(2\pi^2)$:
\begin{equation}
\rho(r) =
\int_{0}^{\infty}\frac{dk}{k}\, \Delta^2(k) \, W(kr) \;  ,
\end{equation}
where $W(kr) \equiv 3 \left(\sin(kr) - kr\cos(kr)\right)/(kr)^3$
is the window function for a top hat (hard sphere).

Thus the reduced second moment measure, $K(r)$, is closely related to
other second--order measures such as the correlation function or the
power spectrum, and it directly measures the mean over--density
of absorbers on scales less than $r$.
However, it has a number of distinct and desirable statistical
properties which have been presented and discussed elsewhere \citep{Quash99,Stein00}.

\subsection{Estimating $K(r)$}

Here we outline our method of estimating $K$, taking into account
all absorber pairs, deferring the complete derivation of the
estimators to the Appendix. We construct two estimators for
$K(r)$, the first, $\hat{K}_\parallel(r)$ (eq. [A7]), using only
absorber pairs on the same line of sight, and the second,
$\hat{K}_\perp(r)$ (eq. [A13]), using only absorber pairs from
different QSO lines of sight.\footnote{Since there are no lines 
of sight in the sample that are within $r_0=0.5$ \hMpc\ of each other,
and since $K(r)$ is a measure of integrated correlation, 
it is not possible to compute $\hat K_\perp(r_0)$ directly, 
using only across--line--of--sight information. However,
we can sensibly consider $\hat K_\perp(r)-\hat K_\perp(r_0)$ for
$r$ greater than the minimum distance $r_0$ between lines of sight (see Appendix).}
Both of these are estimators for the
same reduced second moment measure $K(r)$ defined in equation~(1);
to the extent that these estimators agree, they provide evidence
that any clustering found is not due to absorbers associated with
the QSOs (see \S 1). Furthermore, we can combine the along-- and
across--line--of--sight information to obtain an overall estimator
$\hat K(r)$ (eq.\ [A14]), which should be more accurate than
either $\hat{K}_\parallel(r)$ or $\hat{K}_\perp(r)$,
since we use all of the available information.

In order to estimate $K(r)$ defined in equation~(1), one needs to
estimate the average number of neighbors within a distance $r$ of
a typical absorber. To understand the problems associated with
estimating this quantity, it is helpful to consider the simpler
setting of observing a point process in a single contiguous region
of space. One possible way to estimate $K(r)$ for a point process
observed in a window (or sample region) $A$ with volume $a$ is, 
for each point in $A$, 
count how many other points in $A$ are within $r$ of it, 
sum these counts and then divide by an appropriate quantity that
cancels out the effect of the overall intensity of the process.
\citet{Ripley88} calls such an estimator the ``naive'' estimator.
The problem with this estimator is that it tends to underestimate
$K(r)$ because for a point within distance $r$ of a boundary of
$A$, one may not see all of the points of the process that are
within $r$ of it (see Figure~1). 

There are a number of methods of
correcting for this ``edge effect''\citep{Ripley88,Badd98,Stein93}. 
In this work, we use the
isotropic correction \citep{Ripley88}, which is computationally
well--suited to the setting of a process observed along lines of
sight. To describe this correction, consider the two--dimensional
setting pictured in Figure~1. For a point at $x\in A$, if another
point $y\in A$ is within distance $r$ of $x$, then instead of
giving this event a weight of 1 as in the naive estimator, it is
given weight $w(x,|x-y|)$ equal to the reciprocal of the fraction $\alpha/2\pi$
of the circle of radius $|x-y|$ that is contained within $A$ (see
Figure~1). As with all of the various edge--correction methods, we
then have
\begin{equation}
E\left[ \sum_{x\ne y} 1\{|x-y|\le r\}w(x,|x-y|)\right]=\lambda^2 a K(r) \; ,
\end{equation}
where $1\{.\}$ is the indicator function, which is unity 
if the condition in brackets is true and zero otherwise.
To estimate $K(r)$, we divide this sum by some estimate of $\lambda^2 a$.
Denoting by $N$ the total number of points observed in $A$, we will use
$N(N-1)/a$ as our estimator of $\lambda^2 a$.

Even recognizing that absorbers have a finite volume, 
we get to observe absorbers in
almost {\em none} of the ball of radius $r$ around any absorber,
when estimating the three--dimensional $K$ function.
Thus, whereas with observations in a single contiguous region, $w(x,|x-y|)$
often equals 1
(i.e., when $x$ is not within $|x-y|$ of the boundary of $A$),
for an absorber catalog observed along lines of sight, $w(x,|x-y|)$
will always be much bigger than 1.
The exact form of the weight function is given in the Appendix.
As one should expect, the weight is inversely proportional to the cross section
of the absorbers, or equivalently, to $d^2$.
Fortunately, our estimator for $\lambda^2$
will also be inversely proportional to $d^2$,
so the factor of $d^2$ cancels when estimating $K(r)$.

\citet{Quash99} used what is known as the rigid--motion estimator
to correct for edge effects when estimating
the one--dimensional reduced second moment function.
To apply the rigid--motion method to an estimate that uses across--line--of--sight
information, for every observed distance between pairs
of absorbers less than the maximum distance at which we wish to
estimate $K$, we would have
to apply a three--dimensional rigid motion to the lines of sight,
calculate the amount of overlap between the old and new set of
lines, and then average this amount over all possible directions.
It is not clear how one could do this accurately in practice.
Since we have no evidence regarding which estimator is statistically
superior in the present setting, we use the computationally much simpler
isotropic estimator.

\section{Results}

We have used equation~(A7) and equation~(A14) to estimate the
reduced second moment measure, $K(r)$, for 276 QSO lines of sight,
obtained from the Vanden Berk \etal\ catalog. A total of 345 \CIV\
absorbers have been selected from this heterogeneous catalog,
using selection criteria \citep{Quash96,Quash98} designed to
obtain as homogeneous a data set as possible. We refer the reader
to these papers for a detailed description of the selection
criteria.

Figure~2 shows both $\hat{K}_\parallel(r)$ ({\em dashed line}) and
$\hat{K}(r)$ ({\em solid line}), divided by
their Poisson expectation value, $\case{4}{3}\pi r^3$, for the
entire \CIV\ absorber catalog. These resultant quantities have
expectation value of very nearly unity if there is no clustering
of absorbers (see the Appendix). Figure~2 shows that the two
estimates agree very well (within their estimated errors,
see below) over all distances $r$ from 5 to 300
\hMpc, with $\hat{K}_\parallel(r)$ just slightly larger than
$\hat{K}(r)$ between 30 and 140 \hMpc.

Note that, along the same line of sight,
the number of absorber pairs separated by very large distances is small, 
because of the finite comoving length of the lines of sight (the median
length is 350 \hMpc\ [Quashnock \& Stein 1999]);
thus, in Figure~2, for distances $r$ of 170 \hMpc\ and greater,
$\hat{K}_\parallel(r)$ is noticeably less smooth than $\hat{K}(r)$.
Examination of the numbers of absorber pairs along and across lines of
sight indicates that it is at such distances that the {\em across} line
of sight information dominates the total information available.
In Table~1, we show the number of absorber pairs in the data set, for pairs along and
across lines of sight, as a function of pair separation $r$,
in 10 \hMpc\ bins. We also show the cumulative number of pairs for separations $< r$.
For pair separations $r > 170$ \hMpc, there are more additional
absorber pairs across different lines of sight than along the same line of sight,
whereas for pair separations $r <100$ \hMpc, the opposite is true.
These two numbers delineate the regimes where,
in the Vanden Berk \etal\ catalog, clustering information arises
primarily from pairs across and along lines of sight, respectively.
In particular, Table~1 shows that the sample is too sparse to
significantly compare clustering along and across lines of sight
on scales of less than 100 \hMpc.

If the centers of absorbers form a homogeneous Poisson process in three
dimensions, i.e., if they are unclustered,
then their intersections with the lines of sight form
independent one--dimensional Poisson processes, provided their
size $d$ is sufficiently small compared to the scales of interest (see Appendix).
Thus it is straightforward to simulate the distribution of
$\hat{K}_\parallel(r)$ and $\hat{K}(r)$ under the assumption that
the \CIV\ absorbers are unclustered.

In Figure~3, we show the 95\% region of variation about the
expectation value of unity of, respectively,
$\hat{K}_\parallel(r)$ ({\em dashed line}) and $\hat{K}(r)$ ({\em solid line}), 
divided by their Poisson expectation value, $\case{4}{3}\pi r^3$, for
10,000 simulated data sets of unclustered absorbers with the same
arrangement of lines and average number of absorbers as the Vanden
Berk \etal\ catalog. Their averages ({\em dotted lines}) are very
near the true value of unity, indicating that both estimators are
very nearly unbiased in this case.
We find that the two estimators have essentially the same region
of variation on scales up to $\sim 150$ \hMpc; beyond that, the
range is smaller for $\hat{K}(r)$, reflecting the additional
information contributed by pairs of absorbers across
different lines of sight.

To properly interpret the results in Figure 2, one needs some
measure of the uncertanties in the estimates. \citet{Quash99}
obtained approximate confidence intervals by bootstrapping,
or resampling, lines of sight. 
Such a procedure makes no use of the relative locations
of the lines of sight and is nonsensical when applied to an
estimator using across--line--of--sight information.
A bootstrapping procedure based on resampling regions of the sky
would be more appropriate in the present setting, but the problems
of handling edge effects and the uneven spatial distribution of
lines of sight complicate matters, so it is unclear how well such
a procedure would work. 

To provide a rough idea as to the
uncertainty of our estimators of $K(r)$, we use a crude but simple
approach. We divide the sky into eight regions, each containing
nearly the same total length of lines of sight. We then compute
the sample standard deviation of the eight estimators of $K(r)$ in
the eight regions, assume that the standard deviation $\sigma$ for the
estimator based on all of the data will be smaller by a factor of
$\sqrt{8}$ and then use the overall estimate plus or minus
$2\sigma$ as our confidence interval.

\section{Discussion}
It is clear that Figure~2 shows strong evidence of clustering
on scales up to 100 \hMpc\ ($q_0 = 0.5$), and possibly beyond. 
In addition, $\hat K_\parallel(r)$ and $\hat K(r)$ essentially agree on the magnitude
of the reduced second moment measure up to this distance. 
This is to be expected, since, as is clear from Table~1, there are very
few additional pairs of absorbers coming from different lines of
sight on these scales. Thus, the sample is too sparse to significantly compare
clustering along and across lines of sight on scales of less than 100 \hMpc.

On scales greater than 100 \hMpc, however, there are
significantly more additional pairs of absorbers from different lines of sight,
and it becomes possible to compare their clustering along and across lines of sight.
Since $K(r)$ is an {\em integrated} 
measure of clustering on scales from zero to $r$,
it is necessary to look at estimates of {\em differential} quantities like $K(r_2)-K(r_1)$ 
(with $r_2>r_1$) in order to examine clustering on scales strictly 
{\em between} $r_1$ and $r_2$.
We have investigated the significance of clustering scales greater than 100 \hMpc\
by examining the quantity 
\begin{equation}
\frac{\hat K(r_2)-\hat K(r_1)}{\case{4}{3}\pi(r_2^3-r_1^3)} - 1 =
\frac{\int_{r_1}^{r_2} \xi(u)\, u^2\, du}{\case{1}{3}(r_2^3-r_1^3)}\equiv \xi(r_1,r_2) \; .
\end{equation}
The quantity $\xi(r_1,r_2)$ is the estimated average volume--weighted correlation function on
scales between $r_1$ and $r_2$. 
If the latter two are reasonably close to each other, 
then $\xi(r_1,r_2)$ is an approximate measure of the correlation function $\xi(r)$
on the average scale $r = (r_1+r_2)/2$.
Similarly, we define the quantities
$\xi_\parallel(r_1,r_2)$ and $\xi_\perp(r_1,r_2)$
by replacing $\hat K$ in equation~(5) above by
$\hat K_\parallel$ and $\hat K_\perp$, respectively.
These two quantities are estimates of the average volume--weighted correlation function
that use only along-- or cross--line--of--sight information, respectively.
We have examined clustering on scales $r$ greater than 100 \hMpc\ by 
computing $\xi$, $\xi_\parallel$, and $\xi_\perp$ for a sliding window that is
50 \hMpc\ wide and centered on $r$.

Figure 4 ({\em solid lines}) shows all three quantities,
$\xi(r-25,r+25)$ ({\em top panel}), 
$\xi_\perp(r-25,r+25)$ ({\em middle panel}), 
and $\xi_\parallel(r-25,r+25)$ ({\em lower panel}),
for scales $r$ between 100 \hMpc\ and 300 \hMpc.
Their approximate 95\% regions of variation
(again estimated by dividing the QSO sample into eight subsamples
corresponding to eight different regions of the sky)
are also shown ({\em dashed lines}).
All three curves are quite similar, showing evidence
of clustering on scales $r$ between 100 \hMpc\ and 150 \hMpc,
and all agree with each other within their approximate confidence regions.

What is particularly striking, however, is that $\xi_\perp$
(computed with pairs coming from different lines of sight) essentially
agrees within the errors with $\xi_\parallel$
(computed with pairs coming from the same line of sight);
thus, the additional {\em cross}--line--of--sight information {\em strengthens} the evidence
for clustering on scales from 100 \hMpc\ to 150 \hMpc.
Although the evidence for clustering across lines of sight is,
on its own, only marginally significant, it is consistent with the
amplitude and scale of clustering of absorbers along lines of sight;
indeed, if anything it hints at being even {\em stronger} on these scales.

Such clustering on 100 \hMpc\ to 150 \hMpc\ scales
had been hinted at in the one--dimensional work of
\citet{Quash99}, and has been confirmed by the three-dimensional analysis here.
This argues against claims that all of the apparent
line--of--sight clustering on these scales is due to 
significant contamination along the line of sight by absorbers
that are actually physically associated with the QSO (see \S 1).
Figure~4 shows no evidence on clustering on scales between
150 \hMpc\ and 300 \hMpc, using any of the three estimators;
on scales greater than 150 \hMpc, the absorbers appear to be distributed in
a manner that is consistent with isotropy.
Note that beyond 200 \hMpc, $\xi$ has appreciably smaller estimated variability
than $\xi_\parallel$; this shows how using the full three--dimensional
estimator $\hat K(r)$ can improve the measurement of clustering on very large
scales, even for this modest--sized catalog.

Of course, the lines of sight in the
Vanden Berk et al. absorber catalog are rather sparse, and there
are still only 345 lines that were analyzed here.
The limited size of that catalog, as well as its heterogeneity,
precludes a final, strong statement of the statistical significance
and amplitude of the clustering on scales between 100 \hMpc\ and 150 \hMpc.

As soon as data are available, we will undertake
a new effort at analyzing the clustering
of heavy--element absorbers in the Sloan Digital Sky Survey (hereafter SDSS),
now underway \citep{Margon99}.
The SDSS QSO Absorption--Line Catalog (hereafter the SDSS Catalog)
will include heavy--element
absorption--line systems found in the spectra of about 100,000 QSOs,
with absorbers ranging in redshift from $z=0.5$ to $z\ga 5$.
The SDSS Catalog will be of order 300 times larger than the Vanden Berk \etal\
catalog; furthermore, it will be a homogeneous catalog
with fixed selection and detection criteria for the entire sample.
Also, the density (number of QSOs per solid angle on the sky) of
probing lines of sight will be of order 300 times higher.

Because the SDSS Catalog will have much greater density of lines of sight
than the catalog analyzed here, we should expect a much larger
fraction of the information about clustering of absorbers from
pairs across different lines of sight.
Using simulations,
we have estimated how much smaller the standard errors of the estimators
will be in the 300--times {\em larger} SDSS Catalog;
in particular, we have investigated how
the standard errors will be reduced by using the full three--dimensional
estimator and all the cross--line--of--sight information
in the 300--times {\em denser} SDSS Catalog.

To do this, we have made simulations of randomly placed absorbers
and examined how the standard errors of the estimators change if
the lines of sight are present at an intensity comparable to that
which will be achieved in the SDSS Catalog. We define a region of
space similar to that to be probed by the QSO lines of sight of
the SDSS Catalog:
that section of a cone with half--angle of $45\degr$ and
Earth at its tip, which is bounded by comoving distance $2000 < r
< 3300$ \hMpc\ (corresponding to redshift $1.25 < z < 4$) from
Earth. Lines of sight are placed randomly in this region of space,
with a uniform distribution of comoving lengths between 250 and
450 \hMpc\ similar to that in the Vanden Berk \etal\ catalog
\citep{Quash99}. Mock catalogs were created, with total number,
$m$, of lines of sight equal to 100, 1000, 10,000 and 100,000, and
absorbers randomly placed on all these lines with the same average
number of absorbers per unit comoving length as that observed in
the Vanden Berk \etal\ catalog. 
The catalogs were generated by simulating a
one--dimensional Poisson process on the lines of sight. A set of
10,000 of these unclustered mock catalogs were made for each $m$
except for $m=100{,}000$, for which 100 mock catalogs were made.

The variances of the reduced second moment estimators
$\hat{K}_\parallel(r)$ and $\hat{K}(r)$ (the average, or
expectation value, is, in all cases, very near the true value of
$\frac{4}{3}\pi r^3$) were computed for each $m$, for $5 < r < 300$
\hMpc. We find, not surprisingly, that the standard error of
$\hat{K}_\parallel(r)$ decreases with the total number of lines
$m$ as $1/\sqrt{m}$; however, this reduction in standard error is
{\em constant} over $r$.
For $\hat{K}(r)$, there is an additional reduction for larger $r$.
In Figure~5, we show the ratio of the standard error of
$\hat{K}(r)$ for $m=$ 1000 ({\em short-dashed line}), 10,000 ({\em long-dashed
line}), and 100,000 ({\em solid line}), to that of $\hat{K}(r)$ for
$m=$ 100. As the number of lines of sight increases, there is a
continued reduction in the standard error for larger distances,
due to the additional and relatively more important number of
absorber pairs from across {\em different} lines of sight.

This is displayed more dramatically in Figure~6, where we show the
relative improvement, or ratio, of the standard errors of
$\hat{K}(r)$ to $\hat{K}_\parallel(r)$ for $m=$ 100 ({\em dotted
line}), 1000 ({\em short-dashed line}), 10,000 ({\em long-dashed line}),
and 100,000 ({\em solid line}). With 100,000 lines of sight,
using $\hat{K}(r)$ instead of $\hat{K}_\parallel(r)$ results in an
additional factor of 2 to 20 reduction of the standard error
on scales of 30 to 200 \hMpc, {\em in addition} to the factor of 
$\sqrt{300}$ reduction from the larger sample size,
effectively increasing the sample size 
by an extra factor of 4 to 400 at large distances.

It is the line density (number of lines of sight per solid angle
on the sky, or the number of QSOs per square degree) that
determines the {\em relative} efficiency of $\hat{K}(r)$ to
$\hat{K}_\parallel(r)$, not the total number of lines $m$. This is
also shown in Figure~6, where we show the ratio of the standard
errors of $\hat{K}(r)$ to $\hat{K}_\parallel(r)$, this time for
$m=$ 100 lines of sight, but where the angular density of lines is
10 times ({\em short-dashed and dotted line}) and 100 times  
({\em long-dashed and dotted line}) higher than the initial density;
i.e., the solid angle of the conic region described
above is 10 and 100 times smaller, respectively. Figure~6 shows
that increasing the density of lines by a given factor has the
same effect on the relative efficiency of $\hat{K}(r)$ to
$\hat{K}_\parallel(r)$ as does simply increasing the total number
of lines of sight by that same factor. Of course, the overall size
of the standard errors is governed by the total number of lines $m$ (see Fig. 5).

These comparisons of the errors in $\hat{K}(r)$ and
$\hat{K}_\parallel(r)$ are based on unclustered mock catalogs.
We are investigating, in ongoing simulations, how different the actual relative improvement 
might be in a catalog of clustered absorbers such as the SDSS Catalog.
Using a simple model of voids and clusters \citep{Loh2001}
that mimics the correlation structure of the Vanden Berk et al. catalog, 
we have made 1000 clustered mock catalogs with 100 and 1000 lines of sight.  
We find, for example, that for $r=$ 300 \hMpc, 
the ratio of the standard errors of
$\hat{K}(r)$ to $\hat{K}_\parallel(r)$ is 0.629 and 0.230
for unclustered catalogs with 100 and 1000 lines of sight, respectively (see Fig. 6),
and 0.669 and 0.270 for clustered catalogs with the same numbers of lines.
This indicates that the relative improvement that arises from using
cross--line--of--sight pairs in clustered catalogs is still dramatic and is only slightly
less so (6\% and 17\% change in the above two cases) than for unclustered catalogs. 
Thus, we expect use of the full three--dimensional estimator to substantially
increase clustering sensitivity in the SDSS Catalog, with 
a relative improvement that is only slightly less dramatic than what is shown
in Figure 6 ({\em solid line}). We hope to present more detailed results elsewhere,
with much larger numbers of lines of sight \citep{Loh2001}.

\section{Conclusions}

We present two estimators,
$\hat{K}_\parallel(r)$ and $\hat{K}(r)$, for the
three--dimensional reduced second moment for one--dimensional data (absorber redshifts)
along QSO lines of sight. The first estimator uses absorber pairs along the
{\em same} lines of sight, whereas the latter includes data from across
{\em different} lines of sight.
We apply our algorithm to a sample of 345 \CIV\ absorbers
with median redshift $\langle z\rangle = 2.2$, from the spectra of 276 QSOs,
drawn from the catalog of Vanden Berk \etal.

We confirm the existence of significant clustering of
\CIV\ absorbers on comoving scales up to 100 \hMpc\ ($q_0 = 0.5$),
and find that the additional {\em cross}--line--of--sight
information strengthens the evidence
for clustering on scales from 100 \hMpc\ to 150 \hMpc.
This argues against claims that all the apparent
clustering on these scales is due to 
significant contamination along the line of sight by absorbers
that are actually physically associated with the QSO.
However, the limited size of that catalog, as well as its heterogeneity,
precludes a final, strong statement of the statistical significance
and amplitude of the clustering on scales $\ga 100$ \hMpc.
Also, the sample is too sparse to
significantly compare clustering along and across lines of sight
on scales of less than 100 \hMpc.
There is no evidence of absorber clustering along or across lines of sight for
scales from 150 \hMpc\ to 300 \hMpc.

We show that with a 300--times larger catalog, such as that to be compiled
by the Sloan Digital Sky Survey (100,000 QSOs),
use of the full three--dimensional estimator and cross--line--of--sight information
will {\em substantially} increase clustering sensitivity.
We find that standard errors are reduced by a factor 2 to 20 on scales of 30 to 200 \hMpc,
{\em in addition} to the factor of $\sqrt{300}$ reduction from the larger sample size,
effectively increasing the sample size by an extra factor of 4 to 400 at large distances.
Thus, use of the full three--dimensional reduced second moment estimator will
significantly advance our ability to describe and analyze large--scale clustering
of absorbers, and hence visible matter, from the SDSS Catalog.

\acknowledgments

We wish to acknowledge Don York, Dan Vanden Berk, and all their collaborators
for compiling the extensive catalog of heavy--element absorbers used in this study.
We wish to thank Massimo Mascaro and Ken Wilder for the help they provided with the
computer simulations we made. This work was supported in part
by NASA grant NAG~5-4406 and NSF grant DMS~97-09696 (J.~M.~Q.),
and by NSF grant DMS~99071127 (M.~L.~S. and J.~M.~L.).

\appendix
\section{Three--dimensional reduced second moment
estimators $\hat{K}_\parallel(r)$, $\hat{K}_\perp(r)$ and
$\hat{K}(r)$}

Let $L$ denote the set made up of the $m$ QSO lines of sight, $L_i$ the
$i$th line, $N$ the total number of absorbers, and $\lambda $ the
intensity of the absorber center process, or mean number of absorbers per
unit comoving volume. We use $\partial \mathcal{B}_s(x,u)$ to represent
a shell of inner radius $u$, centered at $x$ and with thickness $s$, $v_R(.)$
to indicate measure in $R$ dimensions and $\# \{ .\} $ the number of
elements in a set. When summing over absorber pairs, we use $\sum
^{\parallel }$ to represent a sum over pairs on the same lines of sight
only, $\sum ^{\perp }$ a sum over pairs across lines of sight only and
$\sum $ a sum over all pairs.

We have to give the absorbers some physical size so that there is a
non--zero probability of intersection between an absorber and the lines of
sight. For simplicity, we assume the absorbers are balls of identical
radius $d$. The clustering we seek to measure is the clustering of the
point process of the centers of absorbers. Although we do not need to
specify $d$, our method requires an approximation that is accurate when
$d$ is much smaller than the distances over which we are interested.
Define $Q = \pi d^2 v_1(L)$, where $v_1(L)$ is the total length of the
lines, so $Q$ is effectively the volume of space within which we can
observe the center of an absorber.

Throughout we assume that $K(r)$ is continuous in $r$. To estimate
$K(r)$, we first estimate $\lambda ^2QK(r)$ and then divide by an
estimate of $\lambda ^2Q$. Estimating $\lambda ^2QK(r)$ involves taking
each absorber in turn and counting the number of absorbers within a
distance $r$. Suppose we have an absorber observed at $x$ on some line
of sight. Let another absorber be observed at $y$ on a possibly different
line of sight, with $|x-y| \le r$. Its center must lie within $d$ of
$y$. There are
generally many other absorbers within $r$ of $x$ that are not observed
simply because they lie too far away from the lines of sight. To take
into account this edge effect each absorber pair $(x, y)$ is given a
weight.

To demonstrate how appropriately chosen weights deal with the problem
of edge effects when estimating $\lambda ^2QK(r)$, we
first show that equation (4) of \S 2.2 holds. 
Define $1_{(0,r]}(u)=1\{0<u\le r\}$ and denote the empty
set by $\phi$.
When observations are in
a contiguous window $A$ with volume $a$ and $\partial
\mathcal{B}_0(x,r) \cap A \ne \phi$ for all $x \in A$,
\begin{eqnarray}
\lefteqn{E \left[ \sum_{x\ne y} 1_{(0,r]}(|x-y|) w(x, |x-y|)\right]} \\
& = &
\lambda ^2 \int_A\!\int_0^\infty\!\int_{\partial
\mathcal{B}_0(0, 1)} 1_{(0,r]}(u)
1_A(x)
1_A(x+(u,\Omega )) w(x, u) u^2 d\Omega\, dK(u)\, dx \\
& = & \lambda ^2 \int_A\int_0^\infty 1_{(0,r]}(u) 1_{A_{u}}(x)
4\pi u^2 dK(u)\, dx \\
& = & \lambda ^2 a \int_0^{r} 4\pi u^2 dK(u) \\
& = & \lambda ^2 a K(r)
\end{eqnarray}
where in equation (A2), 
$(u,\Omega )$ are the spherical coordinates of $y-x$. In
equation (A3), $A_{u}$ is the set of points in $A$ such that $\partial
\mathcal{B}_0(x, u) \cap A \ne \phi $, which is simply $A$ when
$\partial
\mathcal{B}_0(x,u) \cap A \ne \phi$ for all
$x \in A$.

The step from equation (A3) to (A4) holds only for $u$ less than
the circumradius of $A$. \citet{Ohser83} suggested adding a factor
to the estimator so that this step to (A4) is valid
at larger distances. This
factor is simply the ratio of the volumes of $A$ and $A_u$. 
This extension is not of much practical value when $A$ is
a single contiguous region, but it is critical for the
line--of--sight catalog, since
we would otherwise be restricted to estimating $K$ at distances
at most one half the shortest line of sight in the catalog.

Equations (A1)--(A5) demonstrate that estimating 
$\lambda^2 aK(r)$ involves taking shells $\partial
\mathcal{B}_{du}(x, u)$ with $u < r$ for each point of the process $x\in
A$, counting and weighting the number of other points
in these shells and integrating over $u$.
We now seek to mimic this procedure for absorbers observed
along lines of sight.
We first consider the simpler case of $\hat K_\parallel(r)$,
in which only pairs of absorbers along the same line of sight
are counted.
Define $L(x)$ to be the line on which the
absorber at $x$ lies. For each pair $(x, y)$ lying on the same line and
less than $r$ apart, we set
\begin{eqnarray}
w(x, |x-y|) & = & 4d^{-2}|x-y|^2 du/ v_1(\partial
\mathcal{B}_{du}(x,|x-y|) \cap L(x)) \\ & = & 4d^{-2}|x-y|^2 / C_\parallel \nonumber
\end{eqnarray}
where $C_\parallel = C_\parallel(x, |x-y|) = \#\{ \partial
\mathcal{B}_0(x, |x-y|) \cap L(x)\} $. 
In the denominator of equation (A6), $L(x)$ is used and not
$L$, since only absorber pairs on the same line of sight are
considered. Thus $C_\parallel $ takes the value $1$ or $2$. The
estimate of $\lambda ^2QK(r)$ using only absorber pairs on the
same line is then
$$
\sum\nolimits_{x \ne y} ^{\parallel } \frac{1_{(0,r]}(|x-y|)
4|x-y|^2}{d^2C_\parallel} \cdot \frac{Q}{\pi d^2 v_1(A^\parallel_{|x-y|})}
$$
where  $A^\parallel_{|x-y|} = \cup _{k=1}^{m} \{ x
\in L_k : \partial \mathcal{B}_0(x, |x-y|) \cap L_k \ne \phi \} )$
is the subset of $L$ containing points that are a distance $|x-y|$
from at least one other point on the same line. Ohser's extension
is the factor
$\pi d^2v_1(A^\parallel_{|x-y|})/Q$, the proportion of such points in $L$. 
Taking $N(N-1)/Q$ to be the estimate of $\lambda ^2 Q$, we obtain
\begin{eqnarray}
\hat{K}_\parallel(r) & = & \sum\nolimits_{x\ne y} ^{\parallel } \frac{
1_{(0,r]}(|x-y|) 4|x-y|^2}{d^2C_\parallel}\cdot
\frac{Q^2}{N(N-1)\pi d^2v_1(A^\parallel_{|x-y|}) } \nonumber \\ & = &
\sum\nolimits_{x\ne
y}^{\parallel } \frac{ 1_{(0,r]}(|x-y|) 4\pi |x-y|^2 v_1(L)^2
}{C_\parallel N(N-1)v_1(A^\parallel_{|x-y|})} \label{Kparallel}
\end{eqnarray}

We now show that the estimator of $\lambda ^2QK(r)$ using
pairs along the same line of sight has 
an unbiasedness property similar to the one found in (A1)--(A5) as $d\to 0$:
\begin{eqnarray}
\lefteqn{  E\left( \sum\nolimits_{x \ne
y} ^{\parallel } \frac{1_{(0,r]}(|x-y|) 4|x-y|^2}{d^2C_\parallel }
\cdot \frac{Q}{\pi d^2v_1(A^\parallel_{|x-y|})} \right) } \label{unbiased1} \\
 & = &
 \lambda ^2 \int_{{\mathbb R}^3}
\!\int_0^{r}\!\int_{\mathcal{B}_0(0,1)} \frac{1_L(x)1_L(x+(u,\Omega ))
4u^2}{d^2C_\parallel } \cdot \frac{Q}{\pi d^2v_1(A^\parallel_u)}
u^2d\Omega\, dK(u)\, dx  \\
& = &
\lambda^2 Q\int_{{\mathbb R}^3}\!\int_0^{r}
\frac{1_{A^\parallel_u}(x) 4u^2}{d^2C_\parallel} \cdot \frac{\pi
d^2C_\parallel + o(d^2)}{\pi d^2 v_1(A^\parallel_u)} dK(u) dx \label{intoverx} \\
& = &
 \lambda^2Q \int_0^{r} \frac{4u^2}{d^2C_\parallel }(\pi
d^2C_\parallel + o(d^2)) dK(u) \\
& = & \lambda^2 Q K(r)+o(d^2) = \lambda^2 v_1(L) \pi d^2 K(r)+o(d^2). \label{unbiased2}
\end{eqnarray}

Note that when an absorber is observed at $x$ on some line, its
center need not be on the line. In fact, this occurs with
probability zero. All we can infer is that the center is nearby, at most
distance $d$ away. Accordingly, we take $1_{A^\parallel_u}(x)$ to
mean that an absorber center is located so that the center of its
interval of intersection with a line of sight is in
$A_u^\parallel$ , and thus the integral over $x$ in (\ref{intoverx}) yields
$\pi d^2v_1(A_u^\parallel )$ rather than $v_1(A^\parallel_u)$.
Since we do not observe exactly where the centers of the absorbers are,
all of our estimates of $K(r)$ have some small inherent uncertainty that
does not disappear as the size of the observation region
increases. Specifically, as the observation region
grows, our estimates of $K(r)$ converge to some average of
$K(u)$ for $u \in [r-d, r+d]$. 
This does not pose a problem whenever $r$ is much greater than $d$.

We next derive a similar expression for $\hat{K}_\perp(r)$, the
estimator for $K(r)$ using only across--line--of--sight information
(see \S 2.2 for a discussion of its validity). For any absorber
$x$, we consider absorbers $y$ that lie within a distance $r$ on a
{\em different} line of sight. Both the assigned weight and
Ohser's extension have to be changed. 
Define the set $S_x = S_x(|x - y|) =
\partial \mathcal{B}_0(x,|x-y|) \cap (L\backslash L(x))$, the set
of intersections between $\partial \mathcal{B}_0(x,|x-y|)$ and
$L\backslash L(x)$. For the absorber pair $(x, y)$, the assigned weight is
\begin{eqnarray*}
w(x, |x-y|) & = & 4d^{-2}|x-y|^2 du/ v_1(
\partial \mathcal{B}_{du}(x, |x-y|) \cap (L\backslash L(x))) \\
& = & 4d^{-2} |x-y|^2 / C_\perp
\end{eqnarray*}
where $C_\perp = C_\perp(x, |x-y|) = \sum_{s \in S_x}(\cos \theta
_s)^{-1}$ and $\theta _s$ is the angle between the line of sight
on which $s$ lies and the line joining $s$ and $x$. Figure 7
shows why the factor of $(\cos \theta _s)^{-1}$ is used. Referring to
Figure 7, note that although absorber $y$ is observed on line of sight
$k$, the  intersection between $\partial \mathcal{B}_{du}(x, u)$ and
line of sight $i$ is included in the computation of the weight,
since the weight is inversely proportional
to the volume of the region in which the appearance of
an absorber center would yield an observed
location of the absorber in the shell of radius $u$ and
thickness $du$ centered at $x$.
For an estimator that uses only
across--line--of--sight information, the intersection between 
$\partial \mathcal{B}_{du}(x, u)$ and $L(x)$, the line containing $x$,
is not taken into consideration.
The estimate
of $\lambda ^2QK(r)$ is $$
\sum\nolimits_{x,y }^{\perp} \frac{1_{(0,r]}(|x-y|) 4|x-y|^2
}{d^2C_\perp} \cdot \frac{Q}{\pi d^2v_1(A^\perp_{|x-y|})}
$$
where $A^\perp_{|x-y|} = \cup _{k=1}^m \{ x \in L_k :
\partial \mathcal{B}_0(x, |x-y|) \cap (L\backslash L_k) \ne \phi
\}$. Ohser's extension is $\pi d^2v_1(A^\perp_{|x-y|})/Q$ and is the
proportion of points in $L$ that are a distance $|x-y|$ from at
least one other point on another line. This yields as an estimate
of $K(r)$,
\begin{equation}
\hat{K}_\perp(r) = \sum\nolimits_{x, y}^\perp \frac{1_{(0,r]}(|x-y|)
4\pi |x-y|^2v_1(L)^2}{C_\perp N(N-1) v_1(A^\perp_{|x-y|})} \label{Kperp}
\end{equation}
With appropriate changes, i.e. $C_\parallel$ and $A^\parallel_u$
replaced by $C_\perp$ and $A^\perp_u$, steps (\ref{unbiased1})--(\ref{unbiased2}) hold
for $\hat{K}_\perp(r)-\hat{K}_\perp(r_0)$, where $r_0$
is the shortest distance between different lines of sight in
the catalog.

The estimate of $K(r)$ which uses all absorber pairs is now not difficult
to obtain. This estimate is not simply a sum of $\hat{K}_\parallel(r)$ and
$\hat{K}_\perp(r)$.  It is true that it involves a sum of all absorber
pairs, both along and across lines of sight. However, in each term of the
sum, the expression for Ohser's extension and the assigned weight are
different.
For absorber pair $(x, y)$, the probed volume that contributes to the
weight is now a sum of that probed by $L(x)$ and $L\backslash L(x)$:
$w(x, |x-y|) = 4d^{-2}|x-y|^2 / (C_\parallel + C_\perp)$. The
set of points with at least one other point $|x-y|$ away is now a union of
the two sets $A^\parallel_{|x-y|}$ and $A^\perp_{|x-y|}$, which is
$A_{|x-y|}$. With these adjustments, we have

\begin{equation}
\hat{K}(r) = \sum\nolimits _{x\ne y} \frac{1_{(0,r]}(|x-y|) 4\pi
|x-y|^2v_1(L)^2}{(C_\parallel
+ C_\perp)N(N-1)v_1(A^\parallel_{|x-y|} \cup
A^\perp_{|x-y|})}. \label{K}
\end{equation}

Even when using $\hat K(r)$, the fraction of volume of
$\partial \mathcal{B}_{du}(x,u)$
probed by the absorber catalog is 
very small and thus the weights are always much larger than $1$. 
Nevertheless, even in the moderate size catalog used here,
there is sufficient information
to enable us to obtain useful estimates of the
three-dimensional reduced second moment measure at large distances.

\clearpage

\begin{figure}
\epsscale{0.75} 
\plotone{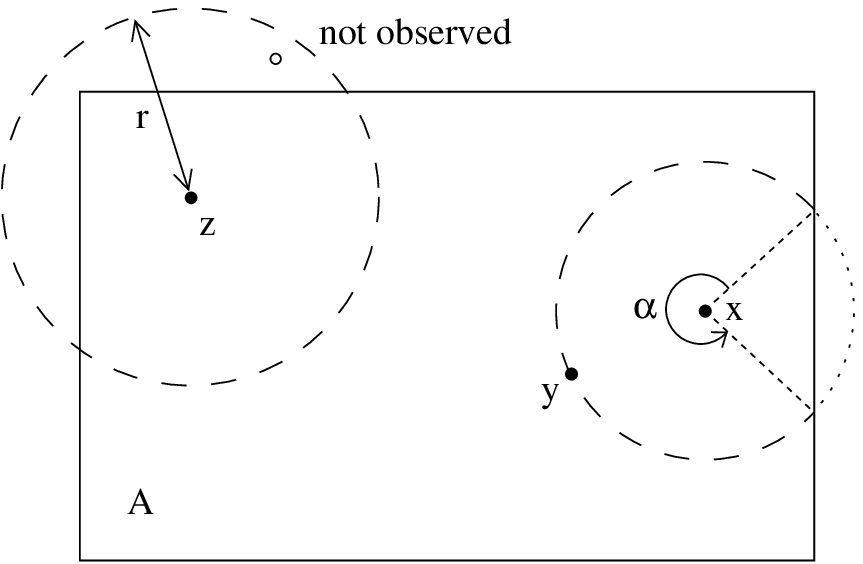} 
\figcaption{Example of an {\em unobserved} 
point ({\em open dot}) within distance $r$ of another
point $z$. Also shown is an {\em observed} point $y$ within distance $r$ of $x$. 
This point is given a weight $w(x, |x-y|)$ equal to the
reciprocal of the fraction $\alpha/2\pi$ of the circle that is contained within
the sample region $A$.}
\end{figure}

\clearpage

\begin{figure}
\begin{center}
\includegraphics[angle=-90,scale=0.7]{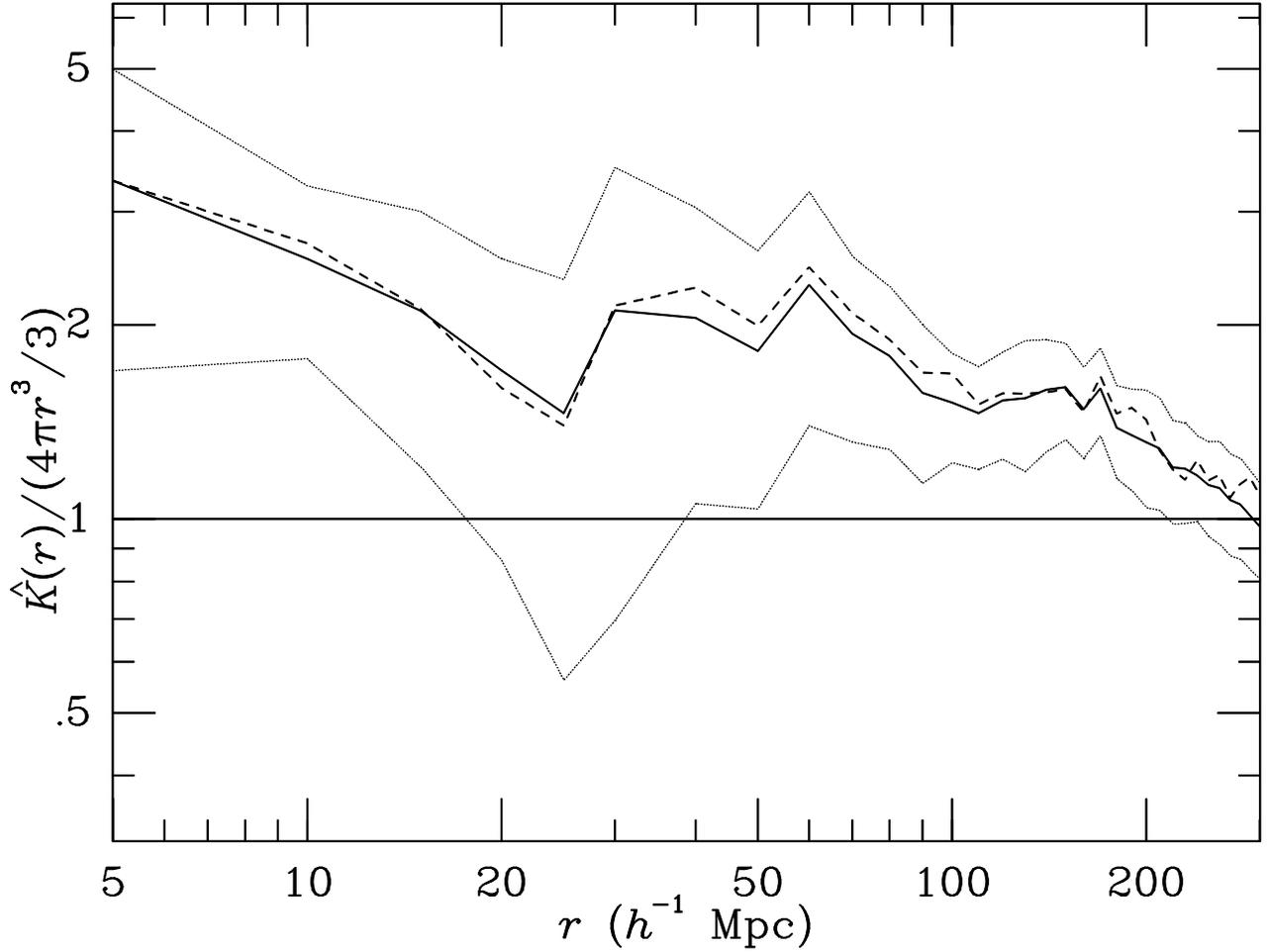}
\end{center}
\figcaption{Estimates of the
reduced second moment measure, $\hat{K}_\parallel(r)$ ({\em dashed line}) 
and $\hat{K}(r)$ ({\em solid line}), divided by their
Poisson expectation $\case{4}{3}\pi r^3$, together with the
latter's approximate 95\% confidence region ({\em dotted line}; see text), 
for the 276 QSO lines of sight, containing a
total of 345 \CIV\ absorbers obtained from the Vanden Berk \etal\ catalog.}
\end{figure}

\clearpage

\begin{figure}
\begin{center}
\includegraphics[angle=-90,scale=0.7]{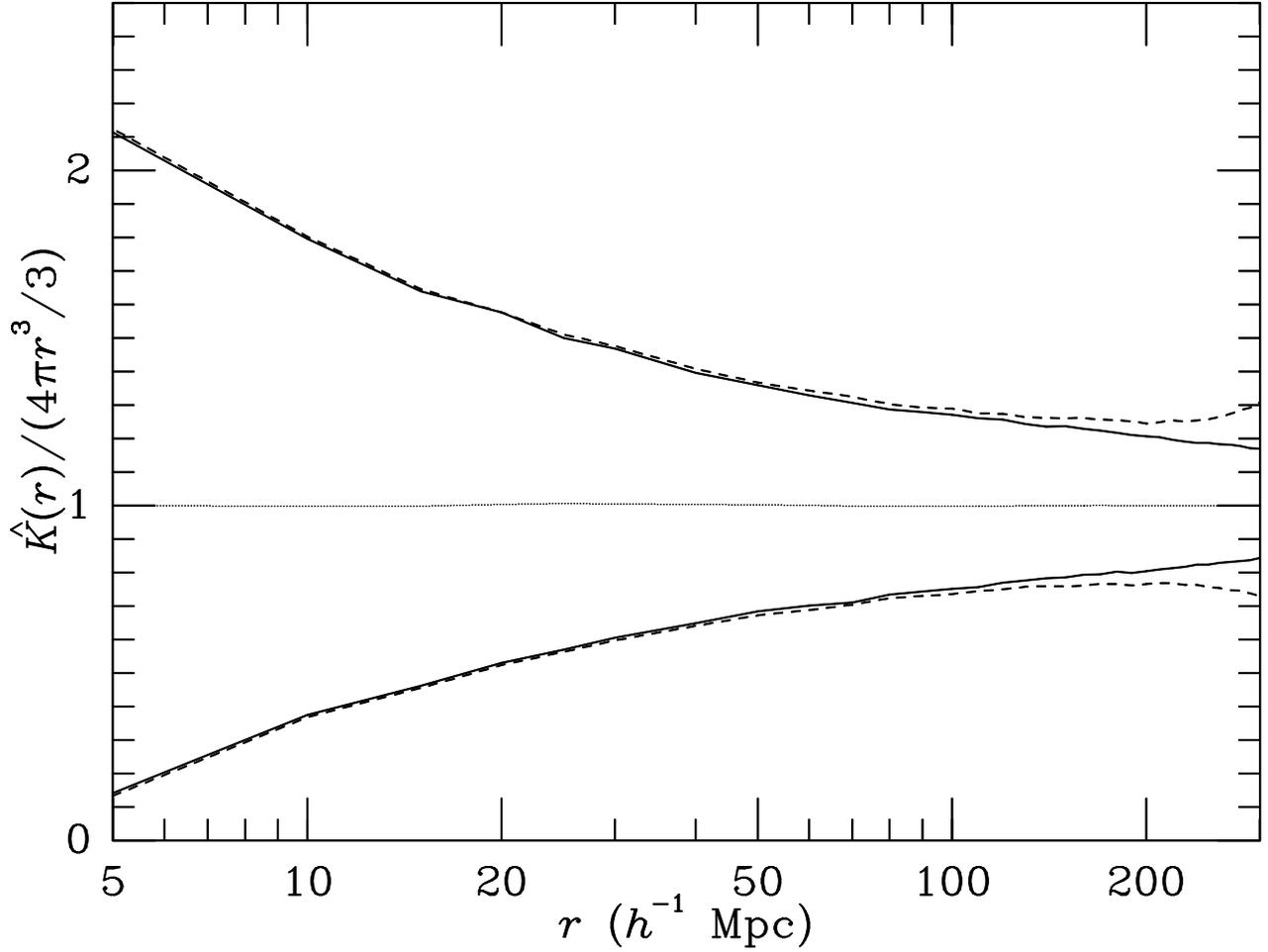}
\end{center}
\figcaption{The 95\% regions of
variation of $\hat K_\parallel(r)$ ({\em dashed line}) and
$\hat{K}(r)$ ({\em solid line}), divided by their Poisson
expectation value, $\case{4}{3}\pi r^3$, for 10,000 simulated data
sets of unclustered absorbers with the same total number of lines
and average number of absorbers as the Vanden Berk \etal\ catalog.
The averages for both estimators ({\em dotted lines}) 
are very near their expectation value of unity.}
\end{figure}

\clearpage

\begin{figure}
\begin{center}
\includegraphics[angle=-90,scale=0.7]{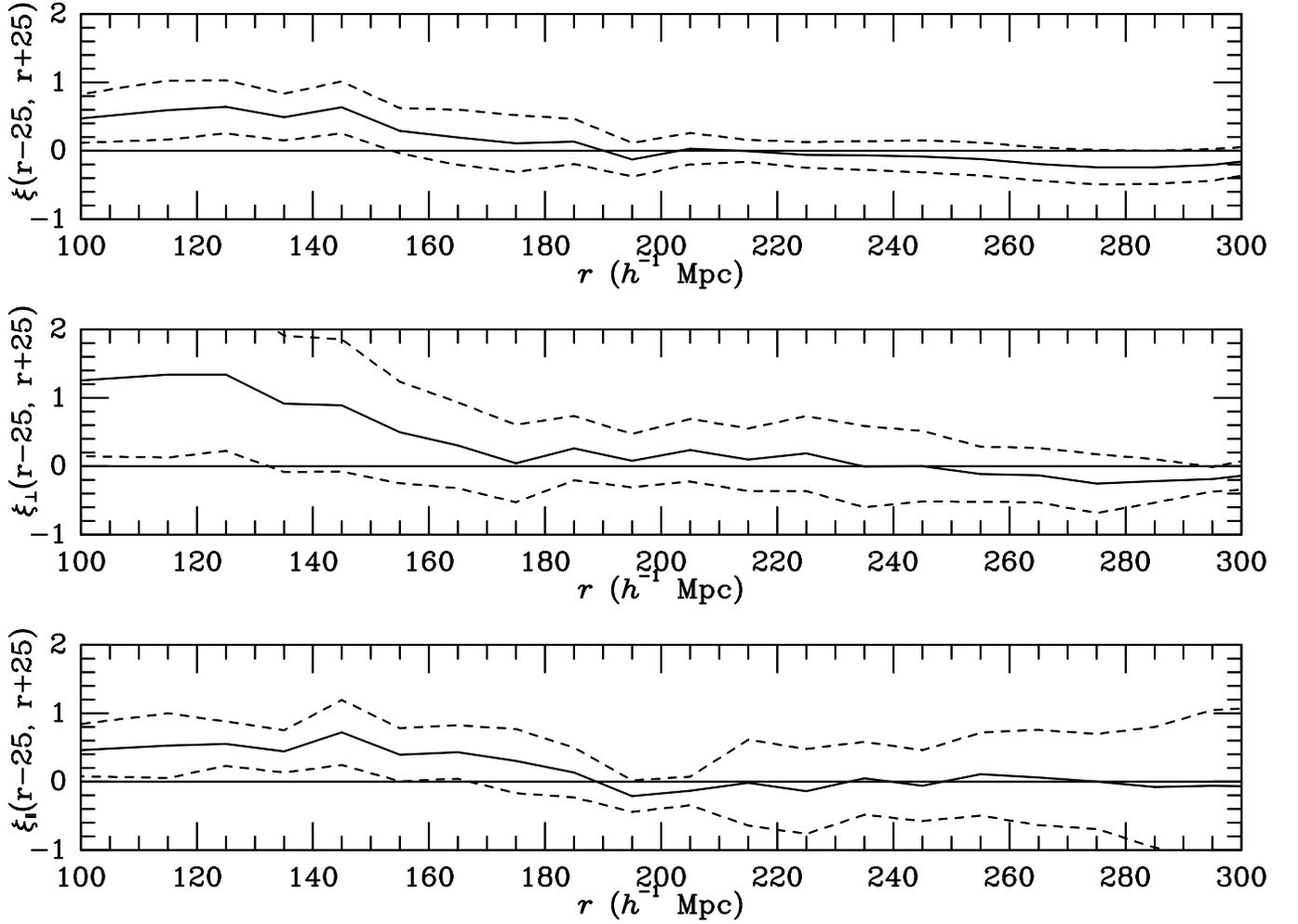}
\end{center}
\figcaption{Average volume--weighted correlation functions ({\em solid lines})
$\xi(r-25,r+25)$ ({\em top panel}), 
$\xi_\perp(r-25,r+25)$ ({\em middle panel}), 
and $\xi_\parallel(r-25,r+25)$ ({\em bottom panel}),
for $100 < r < 300$ \hMpc,
for the 276 QSO lines of sight and 345 \CIV\ absorbers of the Vanden
Berk \etal\ catalog. Their approximate 95\% regions of variation are
also shown ({\em dashed lines}; see text).}
\end{figure}

\clearpage

\begin{figure}
\begin{center}
\includegraphics[angle=-90,scale=0.7]{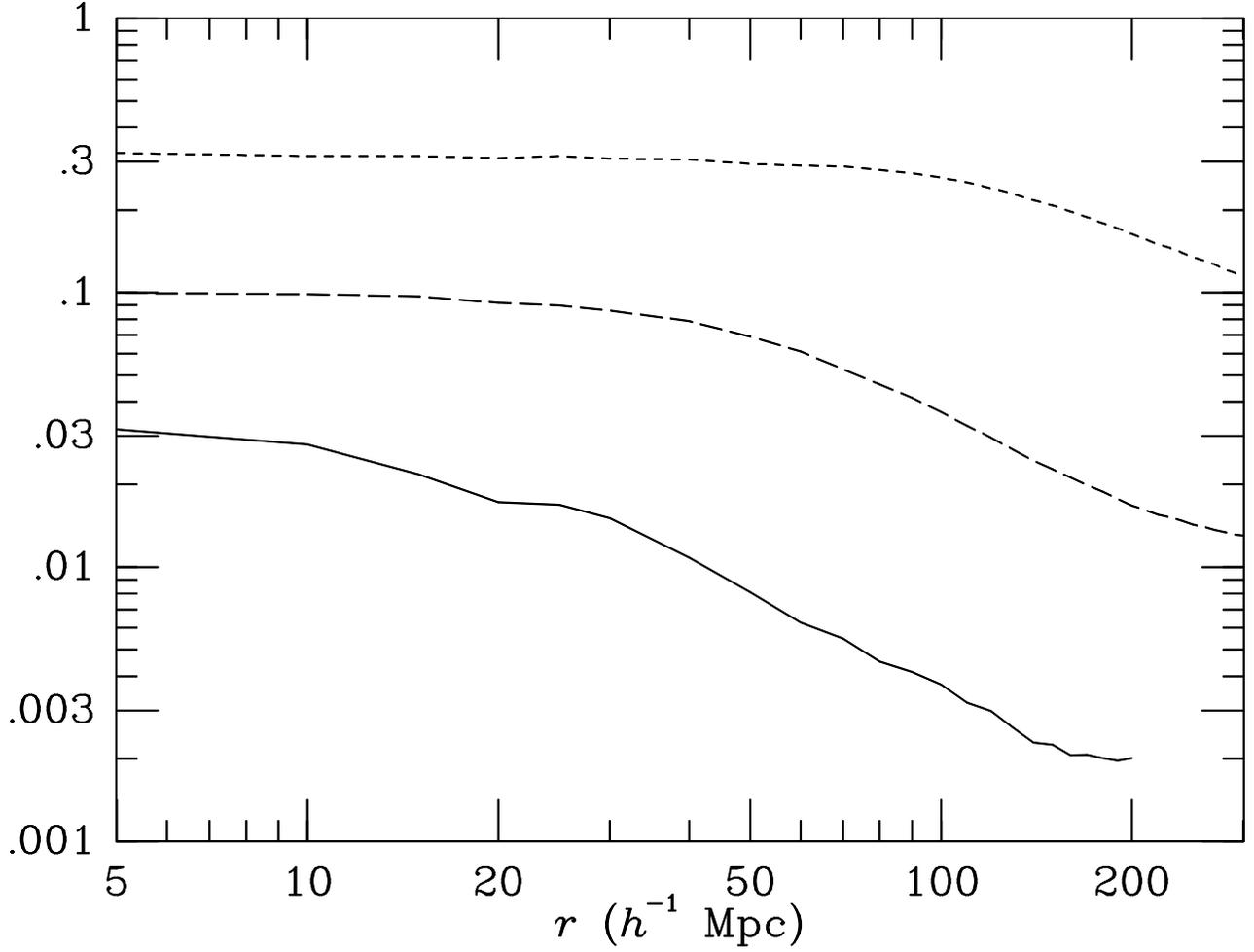}
\end{center}
\figcaption{Ratio of the standard errors of 
$\hat{K}(r)$ for $m=$ 1000 ({\em short-dashed line}),
10,000 ({\em long-dashed line}), and 100,000 ({\em solid line}), to that
of $\hat{K}(r)$ for $m=$ 100, for mock unclustered catalogs. 
Note the continued reduction in the standard
error on larger scales.}
\end{figure}

\clearpage

\begin{figure}
\begin{center}
\includegraphics[angle=-90,scale=0.7]{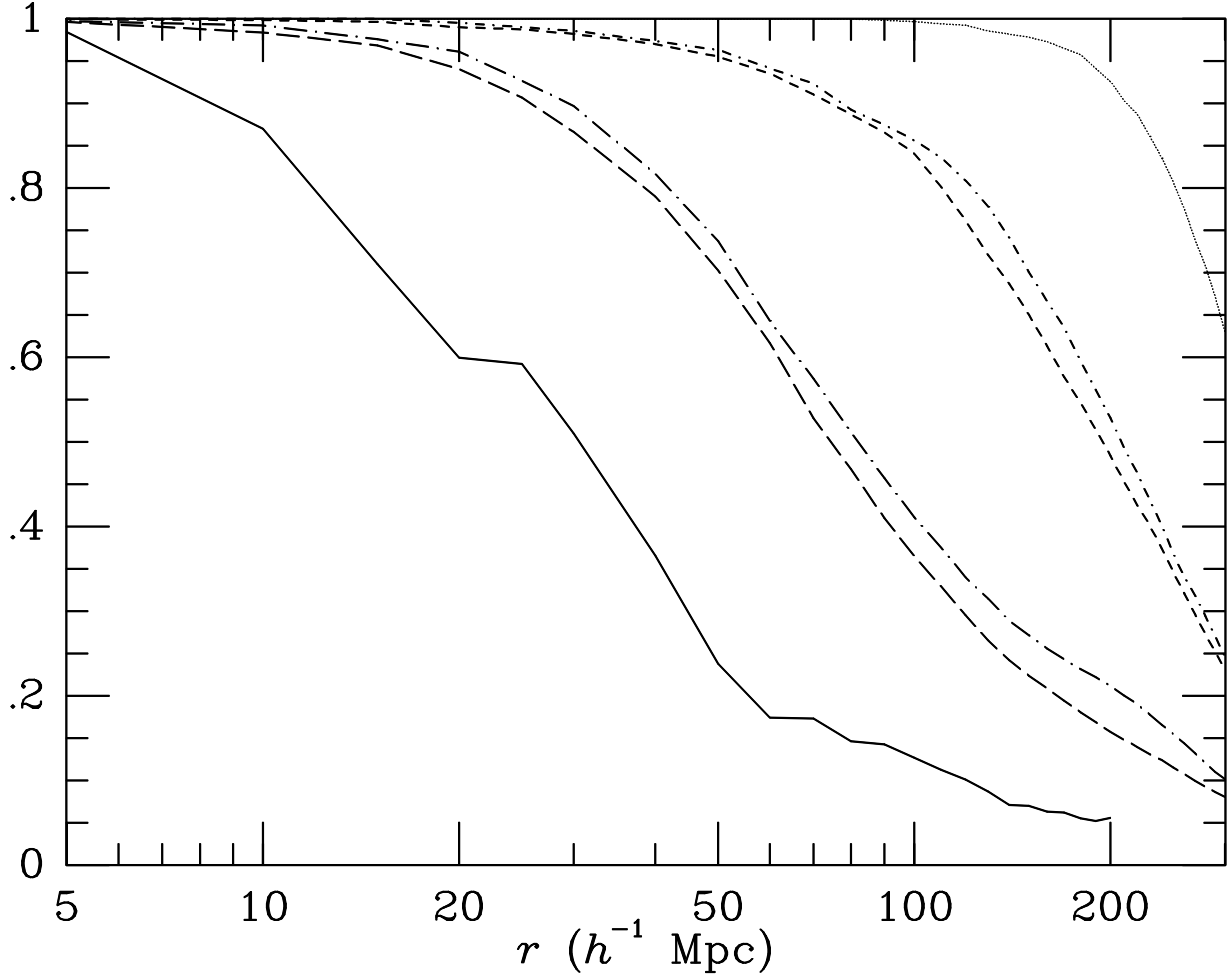}
\end{center}
\figcaption{Ratio of the
standard errors of $\hat{K}(r)$ to $\hat{K}_\parallel(r)$ for $m=$
100 ({\em dotted line}), 1000 ({\em short-dashed line}), 10,000 ({\em
long-dashed line}), and 100,000 ({\em solid line}), with mock
unclustered catalogs. Also shown is the same ratio, for $m=$ 100
lines of sight, but where the angular density of lines is 10 times
higher ({\em short-dashed and dotted line}) and 100 times higher ({\em long-dashed 
and dotted line}).}
\end{figure}

\clearpage

\begin{figure}
\epsscale{0.75}
\plotone{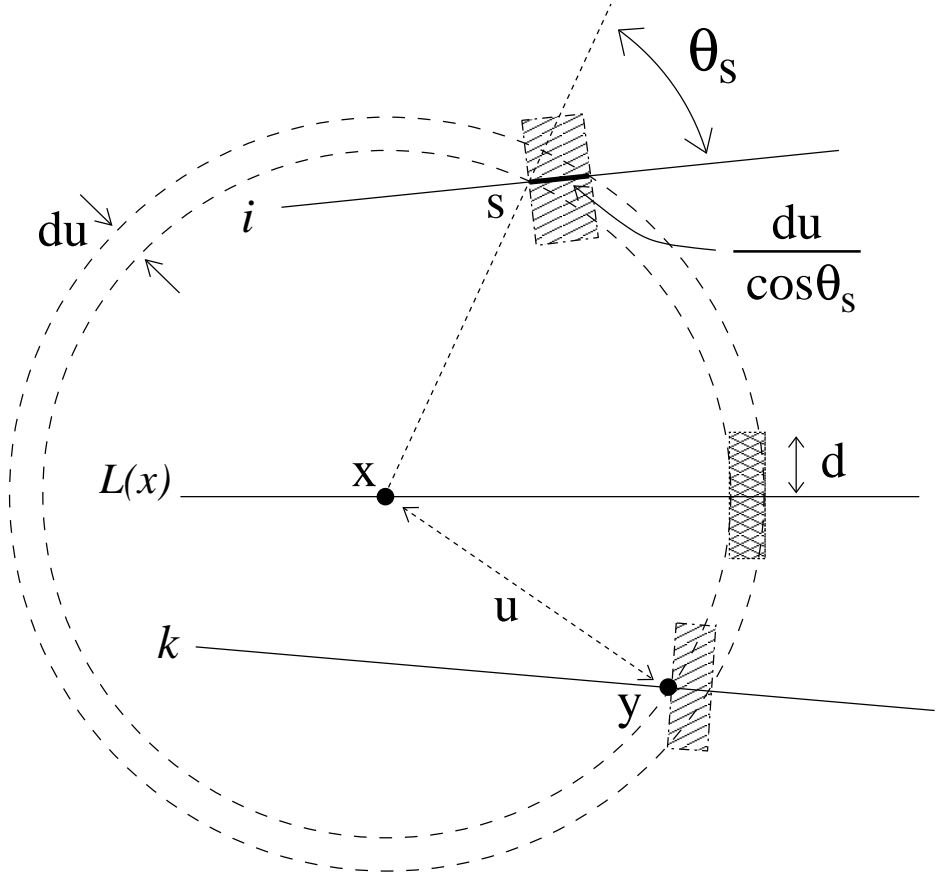}
\figcaption{Schematic showing how to obtain the 
weights for $\hat K_\perp(r)$ and $\hat K(r)$ in a two--dimensional setting. 
Absorbers are observed at $x$ and $y$, and are separated by distance $u=|x-y|\le r$; 
$d$ is the assumed radius of all absorbers.
For $\hat K_\perp(r)$, the weight $w(x,y)$ is inversely proportional
to the sum of the volumes of the two singly hatched regions.
Note that the thickness of the shaded region through
line of sight $i$ is $dr/\cos\theta _s$. 
For $\hat K(r)$, the weight $w(x,y)$ is inversely proportional
to the sum of the volumes of the three hatched regions.}
\end{figure}

\clearpage

\begin{deluxetable}{rrrrrr}
\tablecolumns{6}
\tablewidth{0pc}
\tablecaption{Number of absorber pairs as a function of pair separation $r$}
\tablehead{
\colhead{}    &  \multicolumn{2}{c}{Number in bin} &
\colhead{}    &  \multicolumn{2}{c}{Cumulative Number} \\
\cline{2-3} \cline{5-6} \\
\colhead{$r$ (\hMpc)} & \colhead{along lines}   & \colhead{across lines} &
\colhead{}    & \colhead{along lines}   & \colhead{across lines}}
\startdata
 10 & 63 &  0 & &  63 &   0 \\
 20 & 21 &  8 & &  84 &   8 \\
 30 & 26 &  5 & & 110 &  13 \\
 40 & 30 &  3 & & 140 &  16 \\
 50 & 22 &  5 & & 162 &  21 \\
 60 & 31 &  3 & & 193 &  24 \\
 70 & 17 &  5 & & 210 &  29 \\
 80 & 17 &  7 & & 227 &  36 \\
 90 & 11 &  3 & & 238 &  39 \\
100 & 15 &  6 & & 253 &  45 \\
110 &  9 & 16 & & 262 &  61 \\
120 & 16 &  9 & & 278 &  70 \\
130 & 12 & 10 & & 290 &  80 \\
140 & 14 &  8 & & 304 &  88 \\
150 & 13 &  9 & & 317 &  97 \\
160 &  6 &  7 & & 323 & 104 \\
170 & 18 & 11 & & 341 & 115 \\
180 &  2 & 15 & & 343 & 130 \\
190 & 10 &  9 & & 353 & 139 \\
200 &  6 & 13 & & 359 & 152 \\
210 &  2 & 16 & & 361 & 168 \\
220 &  3 & 19 & & 364 & 187 \\
230 &  4 & 15 & & 368 & 202 \\
240 &  8 &  9 & & 376 & 211 \\
250 &  2 & 10 & & 378 & 221 \\
260 &  5 &  9 & & 383 & 230 \\
270 &  1 & 12 & & 384 & 242 \\
280 &  5 & 17 & & 389 & 259 \\
290 &  4 & 18 & & 393 & 277 \\
300 &  1 & 16 & & 394 & 293 \\
\enddata
\end{deluxetable}

\end{document}